\documentclass[twocolumn,amsmath,floatfix,prb,aps]{revtex4}

\usepackage{color}
\usepackage{bbold}
\usepackage{amsmath}
\usepackage{pifont}   
\usepackage{graphicx} 
\usepackage{dcolumn}  
\usepackage{bm}       
\usepackage{amsfonts} 
\usepackage{amssymb}  
\usepackage{multirow} 
\usepackage{bbm}
\usepackage{hyperref} 
\usepackage{subfigure}
\usepackage{epstopdf}
\usepackage{siunitx}
\usepackage{placeins}

\begin{document}

\newcommand{\ba}{{\bf a}}
\newcommand{\BB}{{\bf b}}
\newcommand{\bd}{{\bf d}}
\newcommand{\br}{{\bf r}}
\newcommand{\bp}{{\bf p}}
\newcommand{\bk}{{\bf k}}
\newcommand{\bg}{{\bf g}}
\newcommand{\bt}{{\bf t}}
\newcommand{\bu}{{\bf u}}
\newcommand{\bq}{{\bf q}}
\newcommand{\bG}{{\bf G}}
\newcommand{\bP}{{\bf P}}
\newcommand{\bJ}{{\bf J}}
\newcommand{\bK}{{\bf K}}
\newcommand{\bL}{{\bf L}}
\newcommand{\bR}{{\bf R}}
\newcommand{\bS}{{\bf S}}
\newcommand{\bT}{{\bf T}}
\newcommand{\bQ}{{\bf Q}}
\newcommand{\bA}{{\bf A}}
\newcommand{\bH}{{\bf H}}
\newcommand{\bX}{{\bf X}}

\newcommand{\bra}[1]{\left\langle #1 \right |}
\newcommand{\ket}[1]{\left| #1 \right\rangle}
\newcommand{\braket}[2]{\left\langle #1 | #2 \right\rangle}
\newcommand{\mel}[3]{\left\langle #1 \left| #2 \right| #3 \right\rangle}

\newcommand{\bdel}{\boldsymbol{\delta}}
\newcommand{\bsig}{\boldsymbol{\sigma}}
\newcommand{\beps}{\boldsymbol{\epsilon}}
\newcommand{\bnu}{\boldsymbol{\nu}}
\newcommand{\bnab}{\boldsymbol{\nabla}}
\newcommand{\bchi}{\boldsymbol{\chi}}
\newcommand{\bGam}{\boldsymbol{\Gamma}}

\newcommand{\bgt}{\hat{\bf g}}

\newcommand{\brh}{\hat{\bf r}}
\newcommand{\bph}{\hat{\bf p}}

\author{M. Fleischmann$^1$}
\author{R. Gupta$^1$}
\author{S. Sharma$^2$}
\author{S. Shallcross$^1$}
\email{sam.shallcross@fau.de}
\affiliation{1 Lehrstuhl f\"ur Theoretische Festk\"orperphysik, Staudstr. 7-B2, 91058 Erlangen, Germany,}
\affiliation{2 Max-Born-Institute for non-linear optics and Short Pulse Spectroscopy, Max-Born Strasse 2A, 12489 Berlin, Germany.}

\title{Moir\'e quantum well states in tiny angle two dimensional semi-conductors}
\date{\today}


\begin{abstract}

The valence band edge in tiny angle twist bilayers of MoS$_2$ and phosphorene is shown to consist of highly localized energy levels created by a ``moir\'e quantum well'', i.e. trapped by the interlayer moir\'e potential. These approximately uniformly spaced energy levels exhibit a richly modulated charge density, becoming ultra-localized at the valence band maximum. The number and spacing  of such levels is controllable by the twist angle and interlayer interaction strength, suggesting the possibility of ``moir\'e engineering'' ordered arrays of quantum dots in 2d twist semi-conductors.

\end{abstract}

\maketitle


\emph{Introduction}: ``Moir\'e materials'', created by the small angle twist of a bilayer material, are an emerging family of two dimensional systems with a remarkable physics of localization. The prototypical moir\'e material, twist bilayer graphene\cite{shall08,shall10,mel11,bist11,lop12,shall13,shall16}, exhibits at the single-particle level strong localization on the moir\'e and, at certain ``magic angles'', low energy bands of vanishing velocity\cite{bist11,shall16}. This, in turn, leads to a rich physics of correlation including Mott states and signatures of high temperature superconductivity, recently observed in experiment\cite{Cao2018,Cao2018a} and currently the subject of intense theoretical investigation. State of the art fabrication techniques now allow for the creation of angle controllable twist bilayers of any 2d material, either semi-metallic or semiconducting. As scattering potentials in semiconductors generally create states in or at the gap edge, moir\'e materials generated from 2d semi-conductors -- in which single layer states scatter off a ``moir\'e potential'' -- might be expected to display a rich physics of localized moir\'e induced impurity states. Recent theoretical work at small twist angles in phosphorene\cite{kang17} ($\theta > 1.8^\circ$) and MoS$_2$\cite{naik18} ($\theta > 2.5^\circ$), that find localization and flat bands, suggest that this is indeed the case.

A continuum approach allows access to the tiny angle limit of these two semi-conductors ($0.3^\circ < \theta \le 1^\circ$) and, just as in graphene, this regime is found to present an enrichment of the physics at larger angles. The nearly dispersionless bands found for $\theta > 1.8^\circ$ and $\theta > 2.5^\circ$, in phosphorene and MoS$_2$ respectively, are seen to develop into an almost uniformly spaced series of energy levels, well described  by the concept of an emergent ``moir\'e quantum well''. This physics of moir\'e confinement generates intensely localized states, with a rich density modulation whose nodal structure is related the level index $n$. As the energy spacing and number of levels is sensitive to the twist angle and interlayer coupling strength, such systems offer the tantalizing possibility of engineering intrinsic moir\'e quantum wells in 2d semi-conductors.

\emph{Model}: The divergence of the moir\'e length in the limit of $\theta\to0$ renders atomistic methods computationally prohibitive for the tiny angle regime, and makes natural the use of continuum methodologies. The continuum approach was thus adopted at an early stage in the investigation of twist graphene\cite{bist11,lop12,shall16}, and has been employed to study several van der Waals heterostructures\cite{als18} as well as partial dislocation networks in bilayer graphene\cite{kiss15,shall17,shall18a}. A crucial criteria that any continuum method must satisfy is that it reproduces, as a controlled approximation, the results of the underlying tight-binding method it is based on. Here we adopt an approach based on an operator equivalence between the tight binding method and a continuum Hamiltonian $H(\br,\bp)$:

\begin{equation}
 \mel{\Psi^{(n)}_{\bk_1\alpha}}{H_{TB}}{\Psi^{(m)}_{\bk_2\beta}} = \mel{\phi_{\bk_1\alpha}^{(n)
 }}{H(\br,\bp)}{\phi_{\bk_2\beta}^{(m)}}.
 \label{opequiv}
\end{equation}
On the left hand side of this equation are Bloch states from the constituent single layers $\ket{\Psi^{(n)}_{\bk_1\alpha}}$ (with layer index $n$, crystal momentum $\bk$, and a combined atomic index $\alpha$), and on the right hand side their natural continuum counterpart: generalized pseudospinor plane waves $\ket{\phi_{\bk_1\alpha}^{(n)}}$. These are given by a plane wave  phase function $e^{i\bk.\br}$ augmented by a pseudospin vector, which in the case of graphene would be simply $(1,0)$ or $(0,1)$ but for the more general materials we consider here are unit vectors in a larger space comprising all the atomic degrees of freedom, i.e. spin, angular momentum, and sub-lattice indices. As shown in Ref.~\onlinecite{M} a $H(\br,\bp)$ exists that satisfies Eq.~\eqref{opequiv} exactly. The interlayer coupling has the general form

\begin{equation}
\label{intL}
 \left[S(\br,\bp)\right]_{\alpha\beta} = \frac{1}{V_{UC}} \sum_{j} M_{j\alpha\beta} 
  e^{-i\bg^{(m)}_i.\br} \hat{t}_{\alpha\beta}(\bK_0+\bG_j+\bp)
\end{equation}
where $\bp$ is the momentum operator, and the sum is over all reciprocal lattice vectors $\bG_j$ of the unrotated layer. The three factors in the sum encode the high symmetry bilayer, the twist momentum boosts, and the electronic coupling through, respectively, the ``$M$ matrices''

\begin{figure*}[!tbph]
  \centering
  \includegraphics[width=0.9\linewidth]{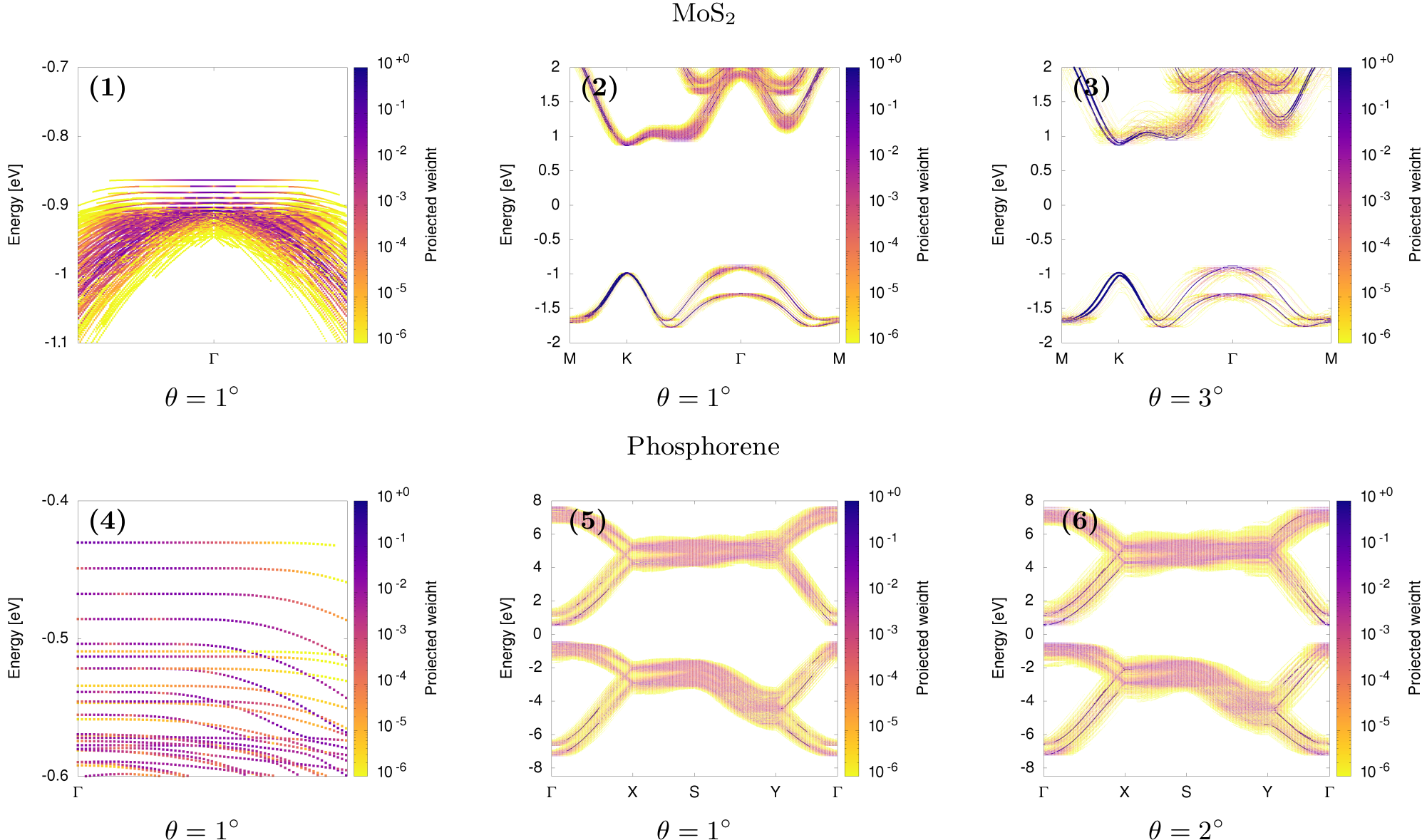}
  \caption{Band structure in extended zone scheme for MoS$_2$ (top row) and phosphorene (bottom row). The weight function is given by the projection of the twist bilayer wavefunction onto the complete set of single layer states at $\bk$; see Eqs.~\eqref{spec} and \eqref{weight}. For the small twist angles shown here the interlayer interaction causes both band broadening, see panels 2,3,5,6, and, near the $\Gamma$ point, the creation of a series of localized levels, see panels 1 and 4.}
  \label{bnd}
\end{figure*}

\begin{equation}
 M_{j\alpha\beta} = e^{i\bG_j.(\bnu_\alpha-\bnu_\beta)}
\end{equation}
the moir\'e momentum $\bg^{(m)}_i = \bG_i - R\bG_i$ ($R$ the rotation operator), and $\hat{t}_{\alpha\beta}(\bq)$ the Fourier transform of the interlayer hopping amplitude between orbitals of atomic indices $\alpha$ and $\beta$. Here the atomic indices represent a compound index of angular momentum, sub-lattice, and spin labels. This approach represents a generalization of that pioneered for graphene\cite{bist11,shall16}, and requires as input only the Slater-Koster tight-binding amplitude functions of a high symmetry bilayer, for which accurate parameterizations exist for both MoS$_2$\cite{cap13} and phosphorene\cite{rud15}.

\emph{Calculation method}: The continuum Hamiltonian expressed in layer space has the form

\begin{equation}
 H = \begin{pmatrix}
      H^{(1)} & S(\br,\bp) \\
      S(\br,\bp)^\dagger & H^{(2)}
     \end{pmatrix}
     \label{1M}
\end{equation}
with $H^{(n)}$ the single layer Hamiltonians, treated exactly at the tight-binding level, and $S(\br,\bp)$ the interlayer coupling operator given by Eq.~\eqref{intL}. To solve the electronic structure problem we use a basis set of combined single layer eigenstates obtained from $H^{(n)}$. In this basis the intra-layer blocks are diagonal, consisting just the single layer eigenvalues, with elements of the inter-layer blocks non-zero only if the difference of the single layer crystal momenta of the basis functions falls on the moir\'e momentum lattice $\{\bg^{(m)}_i\}$. For the tiny angle calculations we undertake $0.3^\circ < \theta < 3^\circ$ we find we require up to 5000 single layer states.

\begin{figure*}[!tbph]
  \centering
  \includegraphics[width=0.7\linewidth]{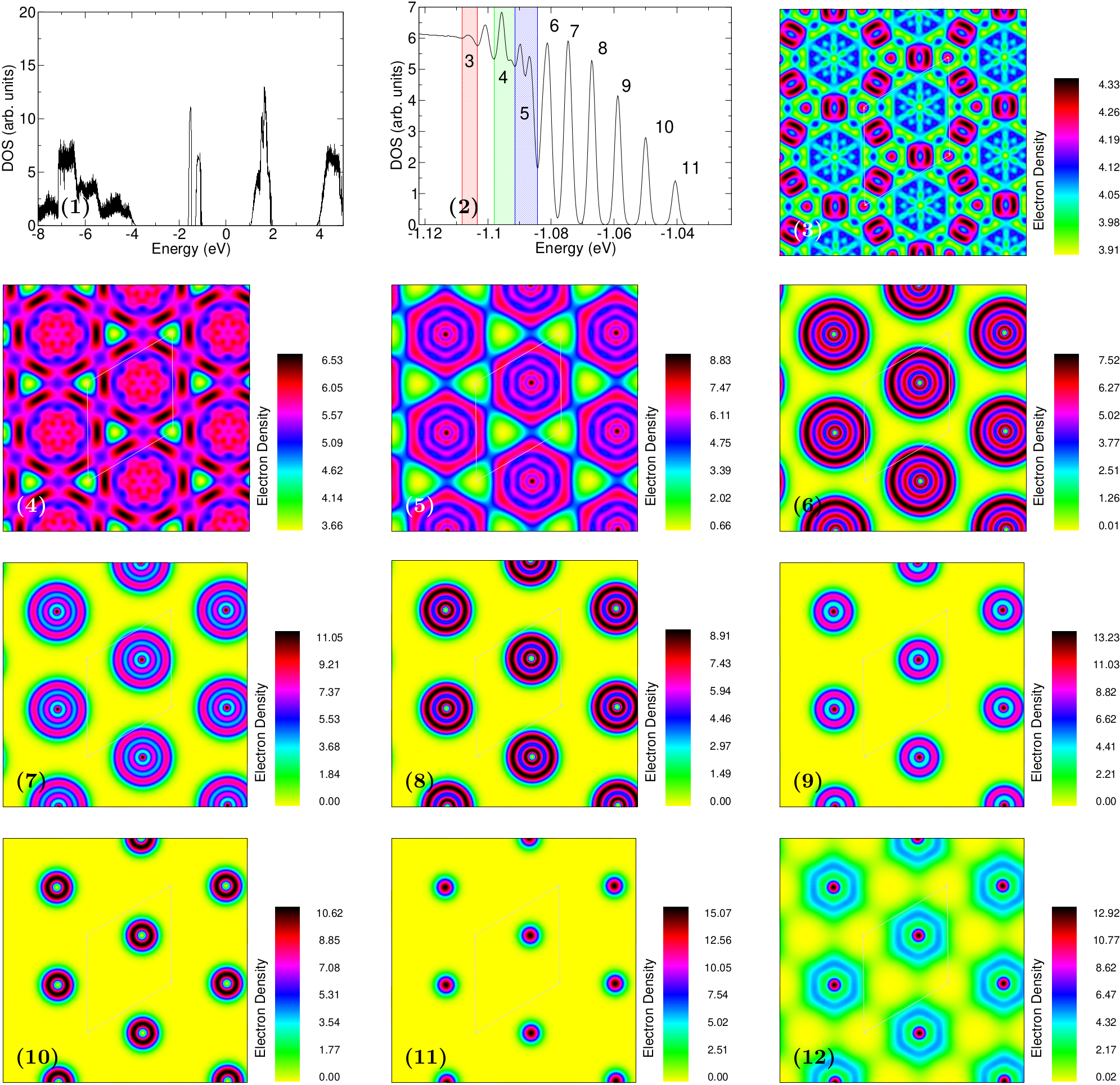}
  \caption{Moir\'e quantum well states in a $\theta=1^\circ$ MoS$_2$ twist bilayer. Panels 1 and 2 show the density of states in a wide energy range and close to the valence band edge respectively. The labels in the latter panel correspond to the integrated charge density shown in panels (3-11); panel 12 exhibits the density for a localized state close to the edge at $\sim 2$~eV.}
  \label{MOS}
\end{figure*}

\begin{figure*}[!tbph]
  \centering
  \includegraphics[width=0.7\linewidth]{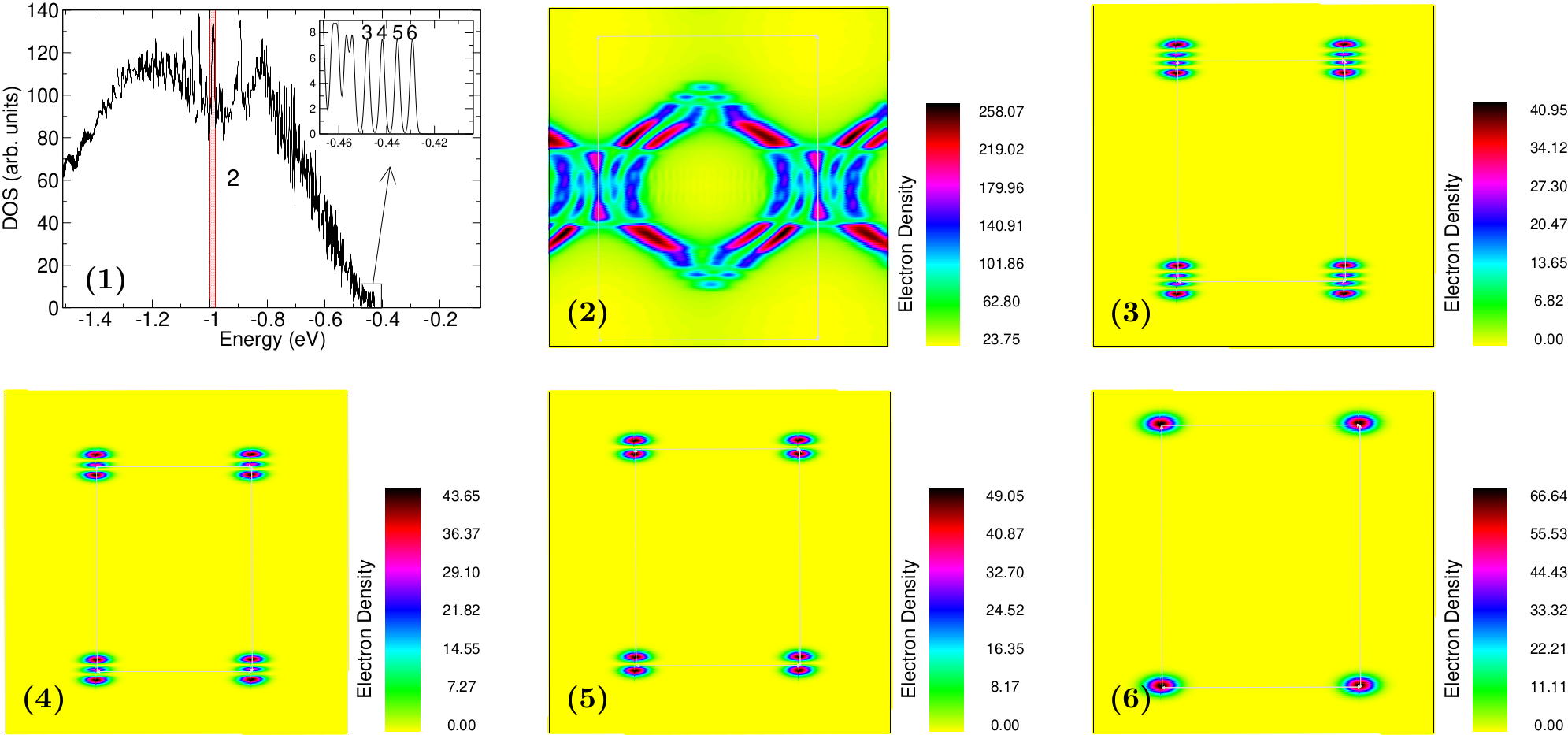}
  \caption{Moir\'e quantum well states in a $\theta=1^\circ$ twist bilayer of phosphorene. In panel (1) is exhibited the density of states of the valence band, with the inset displaying the localized energy levels at the gap edge. Panels (2) and (3-6) show, respectively, the electron density corresponding the the energy window within the valence band, and the edge states indicated in the inset figure.}
  \label{PHOS}
\end{figure*}

In the small twist angle regime it is useful to present the band structure in the extended zone scheme by projection of the twist eigenfunction onto single layer eigenstates, i.e., to plot

\begin{equation}
 \omega(\bk,\epsilon) = \sum_j \rho_{\bk j} \delta(\epsilon-E_{\bk j})
 \label{spec}
\end{equation}
where

\begin{equation}
 \rho_{\bk j} = \sum_{ni} \braket{\psi_{\bk i}^{(n)}}{\Psi_{\bk j}}
 \label{weight}
\end{equation}
is a weight function in which $\ket{\Psi_{\bk j}}$ is an eigenstate of the twist bilayer and $\ket{\psi_{\bk i}^{(n)}}$ a single layer eigenstate from layer $n$ with (single layer) band index $i$. With no interlayer interaction $\rho_{\bk j}=1$ and $\omega(\bk,\epsilon)$ simply represents the superposition of the two single layer band structures. In contrast, in the physical bilayer the interlayer interaction scatters single layer states into other momenta resulting in $\rho_{\bk j} <1$ and an $\omega(\bk,\epsilon)$ describing the formation of mini-bands and mini-gaps in the extended zone.

\emph{Extended zone band structures}: In Fig.~\ref{bnd} we show band structures in the extended zone scheme for MoS$_2$ and phosphorene for $\theta = 1^\circ$. At these twist angles the moir\'e momentum is orders of magnitude smaller than the single layer reciprocal lattice vectors, and the resultant multiple scattering of single layer eigenstates generates an effective broadening of the band structure. In MoS$_2$ this is substantial close to the $\Gamma$ point, but negligible at the valence band K-point, where the twist wavefunctions consist of almost pure single layer states. This reflects both the much weaker coupling at the K point as compared to the $\Gamma$ point, since $|\hat{t}(\bGam)| < |\hat{t}(\bK)|$, but also the absence of states to scatter into near the K valence band edge due to the larger effective mass; note that broadening is seen at K for higher energies (where the effective mass is much reduced), although it is still weaker than that seen close to the $\Gamma$-point. Similarly, in phosphorene the effective mass anisotropy leads to stronger twist induced broadening in the $\Gamma$-X direction of the Brillouin zone as compared to the $\Gamma$-Y direction.

\FloatBarrier

Most interesting, however, is the low energy sector near the $\Gamma$-point, see panels 3 and 6 of Fig.~\ref{bnd}. A series of uniformly spaced and almost dispersionless gap edge states can be seen, which have non-zero band velocity only as their projection weight falls to zero. 

\emph{Moir\'e quantum well states}: A natural explanation for such dispersionless states is that they represent the effectively isolated energy levels of a quantum well created by the moir\'e lattice. If this is so, one would expect (i) level wavefunctions to be isolated on a specific region of the moir\'e and (ii) a structure to the wavefunctions connected to a quantum number $n$ labelling the energy levels, just as occurs in any truly isolated quantum well. As we show in Figs.~\ref{MOS} and \ref{PHOS}, that display the density of states and integrated electron densities of $1^\circ$ twist bilayers for MoS$_2$ and phosphorene respectively, this is indeed the case. 

In panels 6-11 of Fig.~\ref{MOS} the integrated density of the gap edge states in MoS$_2$ is shown corresponding to the edge states labelled in panel 2, revealing both complete isolation from other regions of the moir\'e, as well a level density that is circularly symmetric with $n/2$ periods in the radial density. (The highest energy level has index $n=1$.) As can be seen, the level density become increasing localized towards the valence band maximum. Correspondingly, as the levels join the continuum scattering occurs between quantum wells resulting in a $C_6$ modulation of the level density by the moir\'e lattice and the vanishing of regions of zero charge, panel 5. Into the continuum the density becomes ever more homogeneous, see panels 3 and 4, although remaining weakly modulated by the moir\'e. The quantum well states are localized on the open AA regions of the moir\'e, with the angular momentum character dominated by $p_z$ and $d_{3z^2-r^2}$, just as at the $\Gamma$ point in single layer MoS$_2$.

A similarly structure of the wavefunctions of the energy levels is observed for phosphorene, with the number of maxima in the level density exactly equal to the level index $n$ (with $n=1$ again the highest energy state), see panels 3-6 of Fig.~\ref{PHOS}. The electron density is, however, no longer circularly symmetric as in MoS$_2$, but exhibits a rectangular structure inherited from the rectangular moir\'e of this material, and is localized on the AA' stacked region. We thus conclude that the picture of a gap edge governed by the physics of a moir\'e quantum well provides a good description for both MoS$_2$ and phosphorene. Interestingly, in contrast to MoS$_2$ the electron density in phosphorene is strongly modulation by the moir\'e at all energies. This can be seen in panel 2 of Fig.~\ref{PHOS} in which the integrated density in the window indicated in the density of states plot is shown.

It is worth contrasting the localized and highly structured density of the gap edge states shown in Figs.~\ref{MOS} and \ref{PHOS}, with the well known moir\'e density modulation found in graphene. While semi-classical calculations do lend some credence to the idea of a moir\'e quantum well in graphene\cite{vogl16}, the absence of a gap renders the concept problematic, and indeed the electron density, which is strongly localized on the AA stacked regions, displays neither a structural modulation connected to a level index, or the profound isolation from other regions of the moir\'e seen in Figs.~\ref{MOS} and \ref{PHOS}.

\begin{figure}[!tbph]
  \centering
  \includegraphics[width=0.9\linewidth]{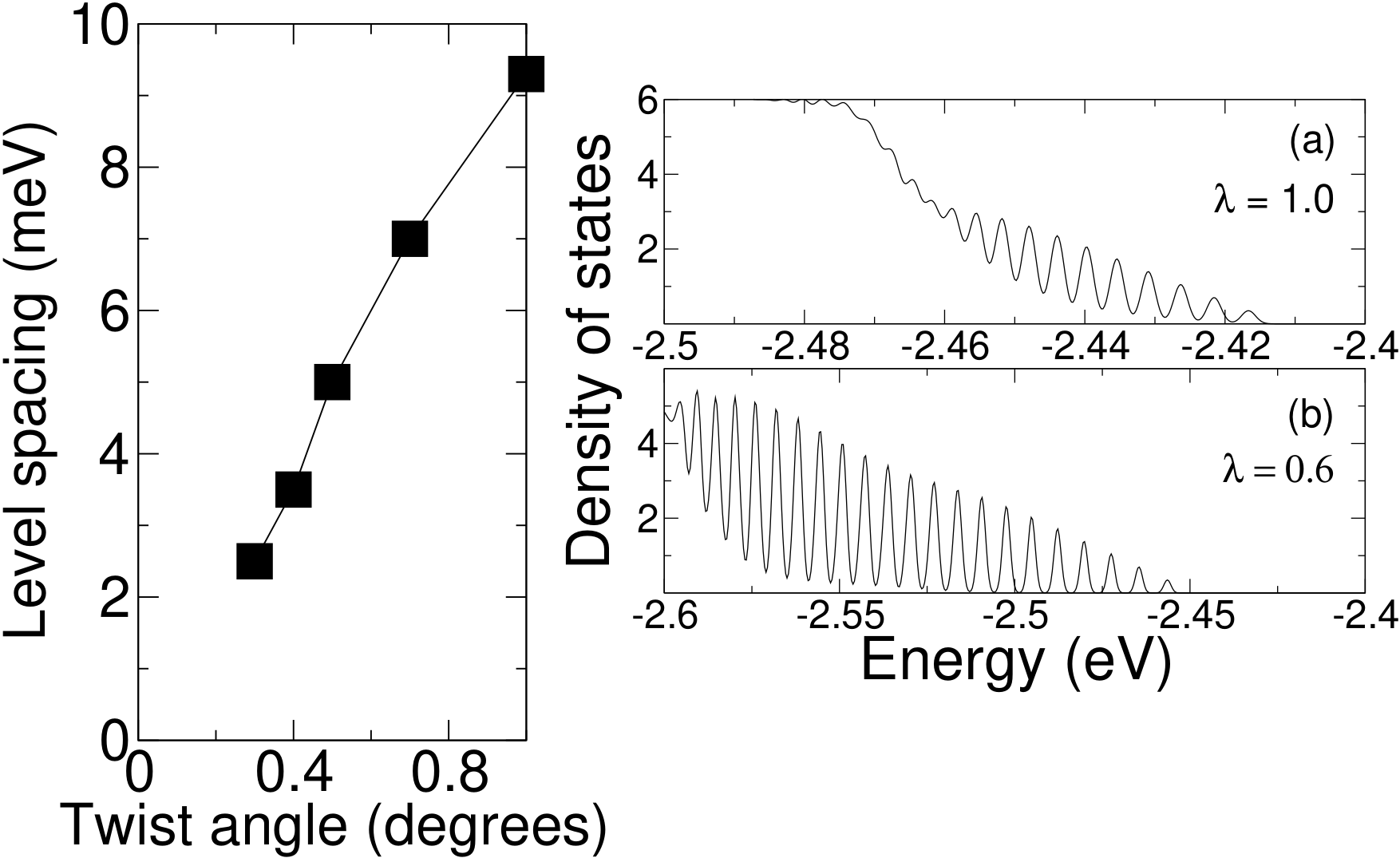}
  \caption{Moir\'e engineering quantum well states in MoS$_2$. In the left hand panel is shown the level spacing of moir\'e quantum well states as a function of twist angle. The right hand panel shows the density of states at the valence band edge for the $\theta=0.5^\circ$ twist bilayer at two interlayer interaction strengths.}
  \label{LEV}
\end{figure}

\emph{Moir\'e quantum well engineering}: The level spectrum of a moir\'e quantum well must, evidently, be controllable through the twist angle. Most obviously, the length scale of the moir\'e well wavefunctions will scale by the moir\'e length. However, the level spacing $\Delta E$ can also be controlled through the twist angle, as shown in Fig.~\ref{LEV}, with $\Delta E$ falling from $\sim 10$~meV to $\sim 2$~meV at $0.3^\circ$ twist angle. Interestingly, the number of levels is quite sensitive to the strength of the interlayer interaction; shown in panel (2) of Fig.~\ref{LEV} are the level states for a $0.5^\circ$ twist bilayer with the interlayer interaction scaled by a factor $\lambda = 0.6$ (21 energy levels) as compared to $\lambda = 1.0$ (10 energy levels).

\emph{Discussion}: We have shown that the $\Gamma$-point valence band edge in frozen moir\'e lattices of tiny angle MoS$_2$ and phosphorene twist bilayers is governed by the physics of a ``moir\'e quantum well'', and features essentially isolated states with wavefunction structure determined by the level index $n$. The $n=1$ state has a spatial extension of the order of a nanometer, comparable to the excitonic radius in single layer MoS$_2$ and thus ordered arrays of quantum wells, created by a tiny twist of a 2d semi-conducting bilayer, form a tantalizing platform for the ``moir\'e quantum well'' engineering of localization and excitonic effects. Spin orbit interaction in MoS$_2$ is likely generate interesting physics in the context of moir\'s quantum well states. While the impact of lattice relaxation in the tiny angle regime will change details of the physics we present here, experience from graphene (where the physics of localization survives relaxation) and first principles calculations at larger angles in MoS$_2$\cite{naik18} and phosphorene\cite{kang17} (in which the creation of flat bands occurs in both ideal and relaxed structures) suggests that the fundamental physics of a moir\'e quantum well we describe will survive relaxation.

\section*{Acknowledgement}

This work was carried out in the framework of SFB 953 of the Deutsche Forschungsgemeinschaft (DFG).


\end{document}